\documentclass[12pt]{article}
\usepackage{epsf}
\def\beq{\begin{equation}}
\def\eeq{\end{equation}}
\def\ap#1#2#3 {Ann. Phys. (NY) {\bf#1} (19#2) #3}
\def\err#1#2#3 {{\it Erratum} {\bf#1} (19#2) #3}
\def\ib#1#2#3 {{\it ibid.} {\bf#1} (19#2) #3}
\def\ijmp#1#2#3 {Int. J. Mod. Phys. {\bf#1} (19#2) #3}
\def\jetp#1#2#3 {JETP Lett. {\bf#1} (19#2) #3}
\def\mpl#1#2#3 {Mod. Phys. Lett. {\bf#1} (19#2) #3}
\def\np#1#2#3 {Nucl. Phys. {\bf#1} (19#2) #3}
\def\pl#1#2#3 {Phys. Lett. {\bf#1} (19#2) #3}
\def\prep#1#2#3 {Phys. Rep. {\bf#1} (19#2) #3}
\def\prev#1#2#3 {Phys. Rev. {\bf#1} (19#2) #3}
\def\prl#1#2#3 {Phys. Rev. Lett. {\bf#1} (19#2) #3}
\def\sjnp#1#2#3 {Sov. J. Nucl. Phys. {\bf#1} (19#2) #3}
\def\spj#1#2#3 {Sov. Phys. JETP {\bf#1} (19#2) #3}
\def\spu#1#2#3 {Sov. Phys. Usp. {\bf#1} (19#2) #3}
\def\zp#1#2#3 {Zeit. Phys. {\bf#1} (19#2) #3}

\begin{document}
\begin{titlepage}
\begin{center}
{\Large \bf Theoretical Physics Institute \\
University of Minnesota \\}  \end{center}
\vspace{0.2in}
\begin{flushright}
TPI-MINN-03/06-T \\
UMN-TH-2131-03 \\
February 2003 \\
\end{flushright}
\vspace{0.3in}
\begin{center}
{\Large \bf  The enhancement of the decay $\Upsilon(1D) \to  \eta \,
\Upsilon(1S)$ by the axial anomaly in QCD
\\}
\vspace{0.2in}
{\bf M.B. Voloshin  \\ }
Theoretical Physics Institute, University of Minnesota, Minneapolis,
MN
55455 \\ and \\
Institute of Theoretical and Experimental Physics, Moscow, 117259
\\[0.2in]
\end{center}

\begin{abstract}
It is shown that the rates of the decays $\Upsilon(1^3D_1) \to  \eta \,
\Upsilon(1S)$ and $\Upsilon(1^3D_2) \to  \eta \, \Upsilon(1S)$ should be
comparable to and likely exceed that of the recently discussed in the
literature two-pion transition $\Upsilon(1D) \to \pi \, \pi \,
\Upsilon(1S)$.  The reason for this behavior is that the discussed
$\eta$ transitions are enhanced by the contribution of the anomaly in
the flavor singlet axial current in QCD.
\end{abstract}

\end{titlepage}

\section{Introduction} The $D$-wave states in the family of the $b \bar
b$ resonances present a new interesting testing ground for the study of
heavy quark dynamics. The $^3D_J$ states with $J=1, 2, 3$ should form a
closely spaced triplet of resonances, of which one (most likely the
$^3D_2$) with the mass of about $10.16 \, GeV$ has been recently
observed\cite{cleo} in the CLEO experiment through the radiative
transitions to and from the $D$ wave state. It is clear however that
similarly to other excited $b \bar b$ resonances there should also be
strong-interaction transitions from the $D$ states to lower resonances
with emission of light mesons, i.e. of two pions, $\eta$, and also a
weaker isospin violating transition with emission of $\pi^0$. In
particular the transitions of the type $\Upsilon(1D) \to \pi\, \pi
\,\Upsilon(1S)$ were discussed in the literature\cite{moxhay,ko,rosner}
in some detail. The interest to the two-pion transitions is explained by
that these are well known to be dominant for the more familiar cases of
the hadronic transitions between $^3S_1$ states, i.e. from $\psi^{'}$ in
charmonium and from $\Upsilon(2S)$, while the rate of similar decays
with the $\eta$ emission being considerably smaller, and the related
isospin violating transition with emission of a single $\pi^0$ being
greatly suppressed further. The amplitudes of these transitions between
the $^3S_1$ states of heavy quarkonium and the pattern of the rates were
understood long ago\cite{vz,ns,is} within the general method of
describing the hadronic transitions in heavy quarkonium using the
multipole expansion\cite{gottfried,mv} in QCD for the interaction of the
heavy quarkonium with soft gluonic field. In the transitions between the
$^3S_1$ heavy resonances the relevant amplitudes for production of the
light mesons are determined\cite{vz} by the low energy theorems arising
from the quantum anomalies in QCD: the emission of an $S$ wave pair of
pions is dominated by the anomaly in the trace of the energy-momentum
tensor\cite{tanom}, while the $P$ wave emission of the $\eta$ is
regulated by the anomaly in the flavor singlet axial
current\cite{aanom}. The anomalous contribution greatly enhances both
rates\cite{vz} in agreement with the available data.

The purpose of the present paper is to point out that the relation
between the rates of the two-pion and $\eta$ transitions from
$\Upsilon(1^3D_{1,2})$ to $\Upsilon(1S)$ should be quite different from
the pattern observed in the transitions between the $^3S_1$ states.
Namely, in the $1D \to 1S$ transitions the pions are emitted in the $D$
wave, and the corresponding amplitude for the production of the pions by
the relevant gluonic operator decouples from the anomaly. On the other
hand, the $P$ wave emission of $\eta$ is still possible for transitions
to the $^3S_1$ state from the $1^3D_J$ resonances with $J=1$ and $J=2$,
and, as will be shown here, is indeed contributed by the axial anomaly
in QCD. The resulting enhancement of the amplitude of the $\eta$
emission turns out to be sufficient to compensate for the suppression
factors inherent in this decay (the flavor SU(3) violation, as well as a
suppression by the inverse of the $b$ quark mass), so that the rate of
the $\eta$ transitions should be comparable to that of the two-pion
ones, and in fact is quite likely to be the largest among the hadronic
transitions from the $^3D_J$ states\footnote{This behavior is
reminiscent of that expected\cite{mv2} for transitions between $^1P_1$
and $^3S_1$ states. There the two-pion emission, also decoupled from the
anomaly, is additionally kinematically suppressed, so that the isospin
violating single $\pi^0$ emission becomes more probable due to the
contribution of the axial anomaly (while the $\eta$ emission is
impossible kinematically).}. A more definite quantitative estimate of
the ratio of the rates, $\Gamma(1^3D_{1,2} \to \eta \, 1^3S_1)/\Gamma(1D
\to \pi \, \pi \, 1S)$, is hindered by the present poor understanding of
a parameter governing the non-anomalous amplitude of production of the
pion pair by gluonic operators, as will be discussed in Sect.4.

Within the QCD multipole expansion treatment of the hadronic transitions
in a heavy quarkonium, outlined below in Sect.2, the evaluation of the
amplitude of the $\eta$ emission requires knowledge of the matrix
element $\langle \eta | G_{\mu \nu} D_\rho G_{\lambda \sigma}| 0
\rangle$, where $G^a_{\mu \nu}$ is the gluonic field tensor, and
$D_\rho$ is the covariant derivative. It will be shown in Sect.3 that
this matrix element is completely determined by the well known
expression\cite{aanom}
\beq
\langle \eta | G^a {\tilde G}^a |0 \rangle = 8 \pi^2 \, \sqrt{2 \over 3}
\, f_\eta \, m_\eta^2~,
\label{eanom}
\eeq
following from the anomaly in the divergence of the flavor singlet axial
current in QCD. In eq.(\ref{eanom}) $f_\eta$ is the $\eta$ `decay
constant', equal to the pion decay constant $f_\pi \approx 130 \, MeV$
in the limit of exact flavor SU(3) symmetry, and $f_\eta$ is likely to
be larger due to effects of the SU(3) violation. Also throughout this
paper the normalization of the gluon field tensor includes the QCD
coupling $g$ (so that e.g. the Lagrangian for the gluon field reads as
$L_g=-G^2/(4 g^2)$).

The amplitudes and probabilities of specific decays are calculated in
Sect.4. Besides a numerical comparison of the rates for the two-pion and
$\eta$ transitions between the $^3D_J$ and the $^3S_1$ resonances in the
$b \bar b$ system, also discussed there are the greatly suppressed
transitions with emission of a $\pi^0$.

Finally, the Section 5 contains a summary of the discussion in the
present paper.

\section{Transition amplitudes in the multipole expansion}

We start with a brief reminder of the leading terms in the multipole
expansion in QCD which are relevant to the discussed
transitions\cite{gottfried,vz}.

The two-pion transition arises in the second order in the $E1$
interaction with the chromoelectric gluon field ${\vec E^a}$ described
by
the Hamiltonian
\beq
H_{E1}=-{1 \over 2} \xi^a \, {\vec r} \cdot {\vec E}^a (0)~,
\label{e1}
\eeq
where $\xi^a=t_1^a-t_2^a$ is the difference of the color generators
acting on the quark and antiquark (e.g. $t_1^a = \lambda^a/2$ with
$\lambda^a$ being the Gell-Mann matrices), and ${\vec r}$ is the vector
for relative position of the quark and the antiquark.

The transitions of the type $^3D_J \to \eta \, ^3S_1$ are induced by the
interference of the $E1$ interaction in eq.(\ref{e1}) with the $M2$ term
containing the chromomagnetic field ${\vec B}^a$ and described by the
Hamiltonian
\beq
H_{M2}=- (4 \, m_Q)^{-1} \, \xi^a \, S_j \, r_i \, \left ( D_i  B_j(0)
\right )^a ~,
\label{m2}
\eeq
where $D$ is the QCD covariant derivative, $m_Q$ is the heavy quark
mass, and ${\vec S}=({\vec \sigma}_1 +{\vec \sigma}_2)/2$ is the
operator of the total spin of the quark-antiquark pair. It should be
noted that the $M1$ term, formally of a lower order in the multipole
expansion, is proportional to the spin-flip operator $({\vec \sigma}_1
-{\vec \sigma}_2)$ and thus does not contribute to transitions between
states with the same total spin.

Using the expressions (\ref{e1}) and (\ref{m2}) the transition
amplitudes are found in the standard way:
\beq
A_{\pi \pi} \equiv A(^3D_J \to \pi \, \pi \, ^3S_1)= \langle \pi  \pi  |
E^a_i E^a_j |0 \rangle \, A_{ij}~,
\label{app}
\eeq
\beq
A_\eta^{(J)} \equiv A(^3D_J \to \eta \, ^3S_1)= m_Q^{-1} \, \left
\langle \eta \left |  E^a_i \, (D_j B_k)^a +  (D_j B_k)^a \, E^a_i
\right | 0 \right \rangle \, A^{(J)}_{ijk}~,
\label{aeta}
\eeq
where $A_{ij}$ and $A^{(J)}_{ijk}$ are the heavy quarkonium amplitudes,
defined as
\beq
A_{ij}={1 \over 32} \, \langle 1S | \xi^a r_i {\cal G} r_j \xi^a | 1D
\rangle
\label{aij}
\eeq
and
\beq
A^{(J)}_{ijk}={1 \over 64} \,  \langle ^3S_1 | \xi^a r_i {\cal G} r_j
\xi^a S_k | ^3D_J \rangle
\label{aijk}
\eeq
with ${\cal G}$ being the Green's function of the heavy quark pair in a
color octet state, and also in these expressions a use is made of the
fact that in matrix elements between color singlet states one can
replace $\xi^a \ldots \xi^b$ by $(\delta^{ab}/8) \, \xi^c \ldots \xi^c$.

The (chromo)electric dipole interaction in eq.(\ref{e1}) does not
involve the spin of the heavy quarks. Thus in the leading
nonrelativistic limit, assumed throughout this paper, where the spin and
coordinate degrees of freedom can be considered as independent, the
amplitude $A_{ij}$ in eq.(\ref{app}) does not depend on the spin
variables of the initial $D$ wave state or of the final $S$ wave one.
For this reason the rates of transitions from each of the $^3D_J$ states
to $\pi \pi \, ^3S_1$ (and also of $^1D_2 \to \pi \pi \, ^1S_0$) are all
the same\cite{yan,moxhay} (modulo small differences in the energy
release, whose effect is formally beyond the assumed approximation) and
can in fact be calculated for spinless quarks.

The amplitude $A^{(J)}_{ijk}$ does depend on the spin-orbital state of
the quark pair and is different for different values of the total
momentum $J$. For the purpose of the present discussion in the leading
approximation of the decoupled spin variables it is convenient to first
represent this amplitude not in the basis of states with definite $J$,
but rather in a form with explicitly factorized spin and orbital
components in the Cartesian coordinates. For this representation we
denote as $\zeta_i$ and $\chi_i$ the spin polarization amplitude of
respectively the initial $^3D$ state and the final $^3S_1$ state, and as
$\psi_{ij}$ the orbital polarization amplitude of the $L=2$ wave in the
initial $D$ states. The tensor $\psi_{ij}$ is symmetric and traceless,
as appropriate for an $L=2$ state, and all these amplitudes are assumed
to be normalized in the standard way, so that the sums over the
polarization states are defined as
\beq
\sum_{pol}\chi_i^* \chi_j = \sum_{pol}\zeta_i^* \zeta_j =
\delta_{ij},~~~\sum_{pol} \psi_{ij}^* \psi_{kl}={1 \over 2} \left (
\delta_{ik} \delta_{jl}+ \delta_{il} \delta_{jk}-{2 \over 3} \,
\delta_{ij} \delta_{kl} \right )~.
\label{norm}
\eeq
In this notation the amplitude $A_{ij}$  (for spinless quarks) is
proportional to $\psi_{ij}$ and can thus be written in terms of a scalar
quantity $A_2$ as $A_{ij}=\psi_{ij} \, A_2$, while the amplitude
$A^{(J)}_{ijk}$ is expressed in terms of the same $A_2$ as
\beq
A^{(J)}_{ijk}={i \over 2} \, \epsilon_{klm} \chi_l^* P^{(J)} \,\psi_{ij}
\, \zeta_m \, A_2~,
\label{ajm}
\eeq
where $P^{(J)}$ is the projector on states with definite $J$, acting on
the product of the spin and orbital polarization amplitudes $\psi_{ij}
\zeta_m$.

The quantity $A_2$ depends on details of dynamics of heavy quarkonium,
and at present is highly model-dependent. For this reason a prediction
of the absolute rates of the discussed decays involves a considerable
uncertainty. Clearly, however, $A_2$ cancels in the considered here
ratio of the rates of the two-pion and $\eta$ transitions, which is thus
determined by the ratio of the matrix elements entering the equations
(\ref{app}) and (\ref{aeta}), describing the production by the gluon
operators of the corresponding light meson states.

\section{Matrix elements for production of light mesons by gluonic
operators}
The gluonic matrix element for the two-pion production in eq.(\ref{app})
multiplies the traceless tensor $\psi_{ij}$ and thus receives no
contribution from the (enhanced) trace anomaly in QCD. Rather this
matrix element is parameterized\cite{ns,moxhay} in terms of the QCD
coupling $\alpha_s$ and the parameter $\rho_G$ introduced in
Ref.\cite{ns} as `the fraction of the pion momentum carried by gluons'.
Using this parameterization, one can write
\beq
A_{\pi^+ \pi^-} = \langle \pi^+ \pi^- | E_i^a E_j^a |0 \rangle \,
\psi_{ij} \, \left ( \chi_k^* \zeta_k \right ) \, A_2=4 \pi \alpha_s
\rho_G \, p^+_i p^-_j \psi_{ij} \, \left ( \chi_k^* \zeta_k \right ) \,
A_2~,
\label{ppp}
\eeq
where the final state with charged pions is assumed for definiteness,
and $p^\pm$ stand for the momenta of the pions in the heavy quarkonium
rest frame.

The matrix element of the gluonic operators in eq.(\ref{aeta}) can be
found as soon as the amplitudes of general Lorentz structure $\langle
\eta |G_{\mu \nu}^a (D_\rho G_{\lambda \sigma})^a|0 \rangle$ and
$\langle \eta |(D_\rho G_{\mu \nu})^a G_{\lambda \sigma}^a|0 \rangle$
are known. These structures can in fact be reduced to (a total
derivative of) the amplitude described by eq.(\ref{eanom}). The
possibility of the reduction of the structures with derivatives to the
expression in eq.(\ref{eanom}) is determined by the general
theory\cite{sv}. Here we present the specific implementation of this
reduction, which uses the following simple algebraic identity valid for
an arbitrary four-vector $p$:
\beq
p_\rho \epsilon_{\mu \nu \lambda \sigma}=p_\lambda \epsilon_{\mu \nu
\rho \sigma}-p_\sigma \epsilon_{\mu \nu \rho \lambda}-p_\mu
\epsilon_{\nu \rho \lambda \sigma}+p_\nu \epsilon_{\mu \rho\lambda
\sigma}~,
\label{eid}
\eeq
where the convention $\epsilon_{0123}=1$ is assumed. The antisymmetry of
the field tensor $G_{\mu \nu}$ then allows one to write the general form
of the first of the discussed matrix elements in the linear order in the
$\eta$ momentum $p$ in terms of two scalars $X$ and $Y$:
\beq
i \,\langle \eta(p) |G_{\mu \nu}^a (D_\rho G_{\lambda \sigma})^a|0
\rangle=X \, p_\rho \epsilon_{\mu \nu \lambda \sigma}+ Y \, \left (
p_\lambda \epsilon_{\mu \nu \rho \sigma}-p_\sigma \epsilon_{\mu \nu \rho
\lambda} \right )~.
\label{gst1}
\eeq
The third structure, allowed by the symmetry and proportional to $(p_\mu
\epsilon_{\nu \rho \lambda \sigma}-p_\nu \epsilon_{\mu \rho\lambda
\sigma})$, is reduced to the first two due to the identity (\ref{eid}).
Furthermore, applying in eq.(\ref{gst1}) the equations of motion (the
Jacobi identity): $D_\rho G_{\lambda \sigma}+D_\sigma G_{\rho \lambda
}+D_\lambda G_{\sigma \rho }=0$, one arrives at the relation $X=2Y$.

Likewise, writing the second of the discussed matrix elements in terms
of two scalars ${\tilde X}$ and ${\tilde Y}$ as
\beq
i \, \langle \eta(p) |(D_\rho G_{\mu \nu})^a G_{\lambda \sigma}^a|0
\rangle={\tilde X} \, p_\rho \epsilon_{\mu \nu \lambda \sigma}+ {\tilde
Y} \, \left ( p_\mu \epsilon_{\lambda \sigma \rho \nu}-p_\nu
\epsilon_{\lambda \sigma \rho \mu} \right )~,
\label{gst2}
\eeq
and applying the Jacobi identity, one finds the relation ${\tilde X}=2
{\tilde Y}$.

The sum of the expressions (\ref{gst1}) and (\ref{gst2}) should combine
into a total derivative, i.e. the sum should be proportional to
$p_\rho$. This
is possible due to the identity (\ref{eid}) under the condition that
${\tilde Y}=Y$, so that all the considered scalar form factors are
expressed in terms of one of them, e.g. in terms of $X$:\,\footnote{An
alternative derivation of two of these relations, namely ${\tilde X}=X$
and ${\tilde Y}=Y$, would be by arguing that in the particular
amplitudes
(\ref{gst1}) and (\ref{gst2}) the operators $G^a$ and $(DG)^a$ can be
considered as commuting with each other, so that the expressions
(\ref{gst1}) and (\ref{gst2}) differ only by re-labeling the indices.}
\beq
{\tilde X}=X,~~~~{\tilde Y}=Y={1 \over 2}\,X~.
\label{xy}
\eeq
Using this relation and contracting the sum of the expressions
(\ref{gst1}) and (\ref{gst2}) with ${1 \over 2}\epsilon^{\mu \nu \lambda
\sigma}$ the form factor $X$ is identified from the equation
(\ref{eanom}) as
\beq
X=-{1 \over 30} \, \langle \eta | G^a {\tilde G}^a |0 \rangle =- {4
\pi^2 \over 15} \, \sqrt{2 \over 3} \, f_\eta \, m_\eta^2~.
\label{xeta}
\eeq

The relations (\ref{gst1}) - (\ref{xy}) fully define the gluonic matrix
element in eq.(\ref{aeta}) in terms of $X$:
\beq
i \, \left \langle \eta \left |  E^a_i \, (D_j B_k)^a +  (D_j B_k)^a \,
E^a_i \right | 0 \right \rangle \ =- 2\, X \, (3 \, p_j \, \delta_{ik} -
p_i \, \delta_{jk})~.
\label{edb}
\eeq
One can now use this expression and the form of the amplitude
$A^{(J)}_{ijk}$ from eq.(\ref{ajm}) to write the full amplitude of the
$\eta$
transition as
\beq
A^{(J)}_\eta= -{2 \over m_Q}\, X \,p_i  \, \epsilon_{jlm} \chi_l^*
P^{(J)} \,\psi_{ij} \, \zeta_m \, A_2~,
\label{aetam}
\eeq
where the symmetry of the tensor amplitude $\psi_{ij}$ is taken into
account.

\section{Relations between the decay rates}
It is quite clear from the equation (\ref{aetam}) that due to the
presence of $\epsilon_{jlm}$ the amplitude of the $\eta$ emission
vanishes for the $J=3$ state, which is totally symmetric in the indices
$i,j,l$, when expressed in terms of the product $\psi_{ij} \zeta_m$.
This previously mentioned behavior is naturally expected, given that the
$\eta$ meson is emitted in the $P$ wave. The projection of the latter
product on the state with $J=1$ is given by
\beq
P^{(1)} \psi_{ij} \zeta_m = {3 \over 10} \, (\delta_{im} \, \psi_{nj}
\zeta_n + \delta_{j m} \, \psi_{ni}\zeta_n-{2 \over 3} \delta_{ij} \,
\psi_{mn} \zeta_{n})~.
\label{j1}
\eeq
The normalization factor here is readily found from eq.(\ref{norm}) and
the condition that the sum over all the polarization states should be
equal to the number of the polarization states of $^3D_1$:
$\sum_{i,j,m} |P^{(1)} \psi_{ij} \zeta_m|^2=3$\,.

Using the explicit expression (\ref{j1}) in the formula (\ref{aetam})
for the transition amplitude, it is a straightforward exercise to find
the square of the amplitude averaged over polarizations of the initial
$^3D_1$ state and summed over the polarizations of the final $^3S_1$
state:
\beq
{\overline {|A^{(1)}_\eta|^2}}={10 \over 9} \, p_\eta^2 {X^2 \over
m_Q^2} \, |A_2|^2~,
\label{p1}
\eeq
where the overline in the l.h.s. denotes the prescribed
averaging-summation operation, and $p_\eta=|{\vec p}|$.

In order to find the similar quantity for the transitions from the
$^3D_2$ states, one generally would have to consider the projector
$P^{(2)}$
similarly to the previous treatment of the projector $P^{(1)}$. However,
at this point it is simpler to use the fact that only the $J=1$ and
$J=2$ states contribute to the `grand total' sum of the square of the
amplitude over all orbital and spin polarization states, and to find the
$J=2$ contribution by using eq.(\ref{p1}) and subtracting the sum over
the $J=1$ states from the `grand total':
\beq
{\overline {|A^{(2)}_\eta|^2}}=
{1 \over 5} \left ( \sum_{J,i,j,k} |A^{(J)}_{ijk}|^2 - 3 \, {\overline
{|A^{(1)}_\eta|^2}} \right ) =
{1 \over 5 } \left ( {40 \over 3} - {10 \over 3} \right ) \, \, p_\eta^2
{X^2 \over m_Q^2} \, |A_2|^2 = 2 \, \, p_\eta^2 {X^2 \over m_Q^2} \,
|A_2|^2~.
\label{p2}
\eeq

A comparison of the latter result with that in eq.(\ref{p1}) immediately
leads to the prediction of the ratio of the decay rates:
\beq
\Gamma(^3D_1 \to \eta \, ^3S_1)= {5 \over 9} \, \Gamma(^3D_2 \to \eta \,
^3S_1)~.
\label{rat12}
\eeq
It should be noted that this relation is obtained in the limit where all
effects of the spin-dependent interaction in the heavy quarkonium are
neglected. The ignored effects include in particular the fine-structure
splitting between the masses of the $^3D_2$ and $^3D_1$. However this
splitting is expected to be quite small, not larger than about $10 \,
MeV$ (for a summary of potential model predictions see e.g.
Ref.\cite{gr}). Thus if the difference in the kinematical factors
$p_\eta^3$ in the decay rate is used as a representative measure of the
contribution of
the unaccounted corrections, one might expect that the accuracy of the
relation (\ref{rat12}) should be about 10\%.

In order to estimate the relative rate of the discussed $\eta$ and $\pi
\pi$ transitions we need to compare the corresponding phase space
integrals of the corresponding amplitude squared at the energy release
$\Delta = M(1^3D_J)-M(\Upsilon) \approx 700\, MeV$, i.e. to compare the
expression
\beq
W_\eta^{(J)}=\int \, {\overline {|A^{(J)}_\eta|^2}} \, 2 \pi \,
\delta(\Delta - \varepsilon_\eta) \, { d^3 p_\eta \over (2\pi)^3 \, 2
\varepsilon_\eta}=
{\overline {|A^{(J)}_\eta|^2}} \, {p_\eta \over 2 \pi}
\label{ephs}
\eeq
(numerically, $p_\eta \approx 435 \, MeV$) for the $\eta$ emission with
the integral
\beq
W_{\pi \pi}=\int {\overline {|A_{\pi \pi}|^2}} \, 2 \pi \,
\delta(\Delta - \varepsilon_1-\varepsilon_2) \,{ d^3 p_1 \over (2\pi)^3
\, 2 \varepsilon_1} \, { d^3 p_2 \over (2\pi)^3 \, 2 \varepsilon_2}
\label{pphs}
\eeq
for the two-pion emission, where in each of the integrals $\varepsilon$
stands for the energy of the corresponding light meson.

The square of the $\pi^+ \pi^-$ transition amplitude from
eq.(\ref{ppp}), averaged over the initial and summed over the final
polarization states, is given by
\beq
{\overline {|A_{\pi^+ \pi^-}|^2}}={8 \over 5} \, \pi^2 \, (\alpha_s
\rho_G)^2 \, \left [ p_1^2 \, p_2^2 +{1 \over 3} ({\vec p}_1 \cdot {\vec
p}_2)^2 \right ] \, |A_2|^2 \to {16 \over 9} \, \pi^2 \, (\alpha_s
\rho_G)^2 \, p_1^2 \, p_2^2  \, |A_2|^2~,
\label{app2}
\eeq
where ${\vec p}_{1,2}$ are the momenta of the pions, and $p_{1,2}=|{\vec
p}_{1,2}|$. Also in the last transition in eq.(\ref{app2}) the averaging
over the relative angle between the momenta is performed, as appropriate
for the purpose of calculating the integral in eq.(\ref{pphs}). A
straightforward integration in eq.(\ref{pphs}) of the expression
(\ref{app2}) yields
\beq
W_{\pi^+ \pi^-}=0.44 \, {\Delta^7 \over 630 \pi} \,  (\alpha_s
\rho_G)^2 |A_2|^2~,
\label{pphsn}
\eeq
where the numerical factor 0.44 accounts for the finite mass of the
pions
in the phase space integral (i.e. for massless pions this factor would
be equal to one). Thus using in eq.(\ref{ephs}) the expression for the
$\eta$ emission amplitude from eq.(\ref{p2}), the ratio of the decay
rates is found as
\begin{eqnarray}
&&{\Gamma(\Upsilon(1^3D_2) \to \eta \, \Upsilon) \over
\Gamma(\Upsilon(1^3D_2) \to \pi^+ \pi^- \Upsilon)} = {W_\eta^{(2)}
\over W_{\pi^+ \pi^-}} = \nonumber \\
&&630 \, {X^2 p_\eta^3 \over 0.44 \, (\alpha_s \rho_G)^2 \, \Delta^7 \,
m_b^2} = {448 \over 15} \, { \pi^4 \over 0.44 \, (\alpha_s \rho_G)^2} \,
{f_\eta^2 \, m_\eta^4 \, p_\eta^3 \over m_b^2 \, \Delta^7} \approx
\left ( {0.64 \over \alpha_s \rho_G} \right )^2~.
\label{repp}
\end{eqnarray}
Here the expression for $X$ given by eq.(\ref{xeta}) is used, and also
in the final numerical calculation rather conservative values of
$f_\eta$ and $m_b$ are assumed: $f_\eta \approx f_\pi \approx 130 \,
MeV$, $m_b \approx 5 \, GeV$.

Clearly, the main uncertainty in evaluation of the ratio of the decay
rates comes from poor knowledge of the dimensionless parameter $\alpha_s
\rho_G$ with both factors normalized at a scale $\mu$ set by the
characteristic size of the quarkonium. The estimates of the relevant
value of this parameter range from $\alpha_s \rho_G \approx
0.2$\cite{moxhay} to $\alpha_s \rho_G \approx 0.59$\cite{ko}, with a
realistic value likely being close to 0.35. In either case, the
numerical result in the equation (\ref{repp}) predicts that the $\eta$
transition rate should be not smaller, but most plausibly larger, than
the rate of the transition with the emission of two pions.

The presented analysis can also be readily applied for an estimate of
the rate of the isospin violating transition with emission of a single
$\pi^0$. In the discussed axial-anomaly-dominated processes the
amplitude of the $\pi^0$ transition is simply related\cite{is} to the
amplitude of the $\eta$ emission by the ratio\cite{aanom} of the
corresponding gluonic matrix elements:
\beq
{A_\pi \over A_\eta}= {\langle \pi^0 | G^a {\tilde G}^a |0 \rangle \over
\langle \eta | G^a {\tilde G}^a |0 \rangle } \, {p_\pi \over p_\eta} =
\sqrt{3} \, {m_d-m_u \over m_d+m_u} \, {f_\pi \over f_\eta} \,{ m_\pi^2
\over m_\eta^2} \,  {p_\pi \over p_\eta}~,
\label{rpea}
\eeq
where $m_u$ and $m_d$ are the masses of the up and down quarks, and the
ratio of the decay rates is thus estimated as
\beq
{\Gamma(\Upsilon(1^3D_J) \to \pi^0 \, \Upsilon) \over
\Gamma(\Upsilon(1^3D_J) \to \eta \, \Upsilon)} = 3 \, \left ( {m_d-m_u
\over m_d+m_u} \right )^2 \, {f_\pi^2 \over f_\eta^2} \, {m_\pi^4 \over
m_\eta^4} \,  {p_\pi^3 \over p_\eta^3}~.
\label{rper}
\eeq
For $(m_d-m_u)/(m_d+m_u) \approx 0.3$\cite{gl} this estimate gives
numerically
about $4 \times 10^{-3}$, so that the rate of the transition with the
emission of a single pion should be quite small.

\section{Summary}

Within the description of hadronic transitions among the levels of a
heavy quarkonium, based on the multipole expansion in QCD, the
interference of the $E1$ interaction in eq.(\ref{e1}) with the $M2$ term
in eq.(\ref{m2}) gives rise to transitions between the $^3D_J$ and
$^3S_1$ resonances in the $b \bar b$ system with the emission of the
$\eta$ meson in $P$ wave. These processes are observable as the decays
$\Upsilon(1^3D_J) \to \eta \, \Upsilon$ of the $D$ wave resonances with
$J=1$ and $J=2$. The dependence on the coordinates of the heavy quarks
of the transition amplitude describing these decays coincides with that
of the amplitude for the transitions $\Upsilon(1^3D_J) \to \pi \pi \,
\Upsilon$. Thus the highly model-dependent factor related to the
dynamics of the heavy quarkonium cancels in the ratio of the amplitudes,
which ratio is thus determined by the amplitudes of production of the
corresponding light meson states by gluonic operators. The relevant
amplitude for the production of $\eta$ (eqs.(\ref{xeta}, \ref{edb})) is
shown to be completely determined by the low energy theorem in
eq.(\ref{eanom}) directly related to the anomaly in the flavor singlet
axial current in QCD. The enhancement of the $\eta$ emission due to the
anomaly contribution, which emission normally would be suppressed by the
inverse of the $b$ quark mass, and the relative smallness of the flavor
SU(3) breaking, makes the rate of this process comparable to, or most
likely greater than that of the two-pion emission, as described by the
equation (\ref{repp}). This pattern of the relative rates is
qualitatively different from that observed in the $2^3S_1 \to 1^3S_1$
transitions, both in charmonium and in the $b \bar b$ system, where the
two-pion transitions dominate. The difference is that in the latter
transitions the two-pion emission amplitude is also enhanced by an
anomaly, in this case the conformal anomaly in QCD, while the amplitudes
of the $D \to \pi \pi  \, S$ decays receive no enhanced anomalous
contribution. The presented treatment, based on the multipole expansion,
predicts the ratio of the $\eta$ transition rates to $\Upsilon$ from the
$^3D_1$ and $^3D_2$ resonances (eq.(\ref{rat12})) with plausibly an
accuracy of about 10\%, while the numerical value of the ratio of these
rates to that of the two-pion emission, given by eq.(\ref{repp}),
contains a considerably larger uncertainty associated with the present
poor knowledge of the parameter $\alpha_s \rho_G$, describing the
emission of two pions (eq.(\ref{ppp})).

\section{Acknowledgements}

\noindent
I gratefully acknowledge enlightening conversations with Arkady
Vainshtein. This work is supported in part by the DOE grant
DE-FG02-94ER40823.

\end{document}